\documentclass[onecollarge]{svjour2}       
\smartqed  
\usepackage{graphicx}
 \journalname{}
\begin{document}

\title{Tunable superconducting properties of a-NbSi thin films and application to detection in astrophysics
\thanks{This work has been partially supported by the ANR (grant No. ANR-06-BLAN-0326), by the Triangle de la Physique (grant No. 2009-019T-TSI2D) and by a Marie Curie Intra European Fellowship within the 7th European Community Framework Programme FP7/2007-2013 (Proposal No. 236122).}}

\author{Olivier Crauste \and Claire A. Marrache-Kikuchi \and Laurent Berg\'{e} \and Sophie Collin \and Youri Dolgorouky \and Stefanos Marnieros \and Claudia Nones \and Louis Dumoulin.}

\institute{O. Crauste \and C.A. Marrache-Kikuchi \and L. Berg\'{e} \and S. Collin \and Y. Dolgorouky \and S. Marnieros \and C. Nones \and L. Dumoulin \at
              Univ Paris-Sud, CSNSM, UMR 8609, Orsay, F-91405, France\\
              \email{claire.marrache@csnsm.in2p3.fr}
\and
              O. Crauste \and C.A. Marrache-Kikuchi \and L. Berg\'{e} \and S. Collin \and Y. Dolgorouky \and S. Marnieros \and C. Nones \and L. Dumoulin\at
              CNRS, Orsay, F-91405, France\\
}
\date{Received: date / Accepted: date}

\maketitle

\begin{abstract}
We report on the superconducting properties of amorphous Nb$_x$Si$_{1-x}$ thin films. The normal-state resistance and critical temperatures can be separately adjusted to suit the desired application. Notably, the relatively low electron-phonon coupling of these films makes them good candidates for an "all electron bolometer" for Cosmological Microwave Background radiation detection. Moreover, this device can be made to suit both high and low impedance readouts.
\end{abstract}

\section{Introduction}
\label{Introduction}

One of the long-standing challenges of cosmology is the comprehension of the forming of the Universe. In this field, the study of the Cosmological Microwave Background (CMB) is a precious source of information. However, the measurement of this primordial radiation is now reaching the sensitivity limit of single pixel bolometer-type detectors. Numerous observations related to the millimetric domain are indeed limited by the photon noise and further improvement of the detectors therefore implies optimizing the focal plane filling factor by the use of large bolometric matrices. Amongst all CMB measurements that the scientific community has yet to perform, the CMB B-mode polarization is probably the most constraining from the instrumental point of view. Detecting the signature of primordial gravitational waves yielding a B-type polarization is one of the utmost goals in today's cosmology and amongst the first objectives in the field.

In order to meet this scientific challenge, the current trend followed by different collaborations is to develop a "bolometric camera" similar to CCDs. Each pixel of this camera will consist in a single detector which sensitivity - measured through its noise equivalent power sensitivity (NEP) - will be at the level of the experimental photon noise : about 10$^{-17}$ W/$\sqrt{Hz}$ for terrestrial use and ten times lower for space-based use. By filling the telescope focal plane with thousands of such pixels, the instrument overall performance will thus be considerably increased.

The base structure of a single pixel for submillimetric measurements requires three components : a radiation absorber, a thermometer and a thermally isolated holder enabling a very weak coupling of the former two with respect to the cold bath. The problem of the thermal decoupling is extremely critical since the sensitivity of a bolometer to an incident power $P$ is limited by its NEP : NEP$^2 = 4k_BT^2G$ and therefore directly depends on the thermal decoupling $G$ between the absorber and the thermometer on the one hand and the cold bath on the other. To date, all devices designed for this application are based on using phonons as vectors for energy transport between the different parts of the bolometer. This implies that the thermal decoupling is achieved through very thin mechanical structures involving delicate clean room processes that are difficult to implement on macroscopic lengths such as those needed for this particular application.

We have recently proposed an innovative solution to this problem \cite{Marnieros2009} where these caveats are lifted : in the envisioned "all electron bolometer", the incoming energy will be directly captured and measured in the sensor electron bath and the thermal decoupling will be ensured via electron-phonon decoupling at low temperature. The idea of using electron-phonon decoupling has been introduced earlier \cite{Gousev1996}, but in our imagined device, the absorption will also take place within the superconducting film, so that complicated antenna designs are no longer needed. One possible device consists in a single superconducting thin film combining the absorber and the thermometer functions. The incoming photons having energies larger than the superconducting gap ($h\nu \gg k_BT_c$) \cite{Lee2000}, they will see the film normal resistance and the electromagnetic wave will be absorbed by the electrons. The radiation-induced electronic temperature increase will be measured by the resistivity variation of the superconducting film, similarly to a Transition-Edge-Sensor \cite{Irwin2005}.  The superconducting film then has to meet four requirements : 1. In order to be suitable for matrix fabrication, it must have a size of the order of the wavelength ($\simeq$ 1 mm); 2. For the wave absorption to be optimum, the film must meet the vacuum impedance (i.e. a sheet resistance of 377 $\Omega$ per square); 3. It must have a competitive electron-phonon decoupling in order to ensure a small enough NEP; 4. Its response time must be suitable for experimental observation (typically, a few ms is acceptable). The first condition is easily met thanks to standard evaporation or lithography techniques. In the present paper, we will focus on the three other pre-requisites.

The present work centers on superconducting amorphous Nb$_x$Si$_{1-x}$ thin films that have already shown interesting properties as thermometric thin films \cite{Dumoulin1993,Hertel1983}. By carefully adjusting the films composition $x$ and thickness $d$, we show that the superconducting properties, namely the normal sheet resistance $R_n$ and the critical temperature $T_c$ can be independently adjusted to build a detector suitable for submillimetric measurements.

\section{Experimental Procedure}
\label{sec:experimental_procedure}

The amorphous NbSi thin films have been synthesized under ultrahigh vacuum (10$^{-8}$ to 10$^{-7}$ mbar) by electron beam (e-beam) co-deposition of Nb and Si onto sapphire substrates coated with a 50-nm-thick SiO underlayer. Each evaporation was controlled in situ
by a dedicated set of piezoelectric quartz in order to precisely monitor the composition and the thickness of the deposition. Its homogeneity throughout the film volume was also carefully checked. The in situ measurements were then systematically and successfully compared to ex situ Rutherford Back Scattering (RBS) measurements. The compositions ranged from 14\% to 18\% and thicknesses $d$ ranging from 45 {\AA} to 1000 {\AA}. The films size was of a few mm$^2$ and the samples have all been annealed at $70\,^{\circ}\mathrm{C}$.

The electrical characteristics of the films have been measured down to below 10 mK using a dilution refrigerator. Resistance measurements were performed using a standard ac lock-in detection technique and all electrical leads were filtered from radio frequency at room temperature.

\section{Results}
\label{sec:Results}

\begin{figure}
  \includegraphics[width=0.8\textwidth]{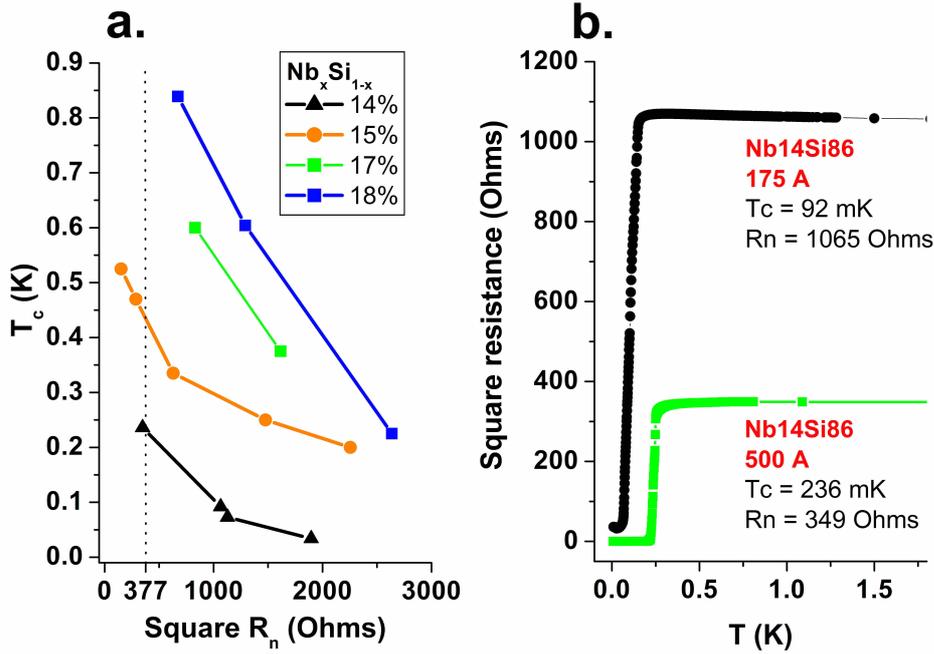}
\caption{a. Superconducting critical temperature as a function of the normal sheet resistance for different compositions. The two parameters can thus be independently tuned. The lines are guides to the eye. b. Resistance characteristics of the 175 {\AA} and 500 {\AA} thick Nb$_{14}$Si$_{86}$ samples.}
\label{fig:1}
\end{figure}

The results for the studied superconducting thin films are summarized in table \ref{tab:1}. The superconducting critical temperature strongly depends on the niobium composition : for the studied samples, it has been varied from 34 mK (for the 105 {\AA} thick Nb$_{14}$Si$_{86}$ sample) to 840 mK (for the 125 {\AA} thick Nb$_{18}$Si$_{82}$ sample). "Thick" films ($d>75$ {\AA}) have a constant resistivity, and hence a normal sheet resistance $R_n$ inversely proportional to the thickness. Thinner films have a resistivity larger than the bulk value due to prominent 2D effects. As can be seen from figure \ref{fig:1}.a., $T_c$ and $R_n$ can thus be separately tuned and adapted to the desired application. Regarding sub-millimetric measurements, the 500 {\AA} thick Nb$_{14}$Si$_{86}$ sample is an optimal candidate for wave absorption, while the 175 {\AA} thick Nb$_{14}$Si$_{86}$ sample has a $T_c$ suitable for an operating temperature of 100 mK (the working temperature for the Planck satellite experiment \cite{Morgante2009}) (figure \ref{fig:1}.b.).

Furthermore, it must be noted that the characteristics are here given for films annealed at $70\,^{\circ}\mathrm{C}$. A heat treatment at higher temperature renders the sample more insulating : the critical temperature is lowered and the normal resistance enhanced. The annealing can therefore be used to fine-tune the desired $T_c$ and film resistance (see for example \cite{Querlioz2005}). It has been shown elsewhere \cite{Crauste2010} that, for a 500 {\AA}-thick Nb$_{13.5}$Si$_{86.5}$ sample, the resistivity and the superconducting critical temperature can be linearly tuned by the annealing temperature. When the annealing temperature is varied from $70\,^{\circ}\mathrm{C}$ to $250\,^{\circ}\mathrm{C}$, the resistivity increases by 17\% while the $T_c$ drops by 61\%. It is therefore conceivable that the annealing be used at the end of the detector fabrication process to fine-tune the adaptation to the operating temperature, without much affecting the normal resistance.

\begin{table}
\caption{Superconducting properties of the Nb$_x$Si$_{1-x}$ samples : composition $x$, thickness $d$, normal sheet resistance $R_n$ (at 4.2 K) and critical temperature $T_c$.}
\label{tab:1}
\begin{tabular}{llll}
\hline\noalign{\smallskip}
$x$ (\%) & $d$ (\AA) & $R_n$ ($\Omega$) & $T_c$ (mK)\\
\noalign{\smallskip}\hline\noalign{\smallskip}
14 & 105 & 1896 &  34\\
14 & 150 & 1128 &  73\\
14 & 175 & 1065 &  92\\
14 & 500 & 349  & 236\\
\noalign{\smallskip}\hline
15 &  75  & 2256 & 200\\
15 & 100  & 1476 & 250\\
15 & 250  &  630 & 335\\
15 & 500  &  287 & 470\\
15 & 1000 &  152 & 525\\
\noalign{\smallskip}\hline
17 & 75  &  1615 & 375\\
17 & 125 &   831 & 600\\
\noalign{\smallskip}\hline
18 & 45  &  2637 & 225\\
18 & 75  &  1291 & 604\\
18 & 125 &  673  & 840\\
\noalign{\smallskip}\hline
\end{tabular}
\end{table}

In order to test the competitiveness of these films as detectors, electron-phonon coupling measurements have been performed using I-V characteristics, following the procedure described in \cite{Marnieros2000}. Note that the films were current-biased, so that no phase separation (such as reported in \cite{Cabrera2000}) was observed. For the considered samples, the electron-phonon coupling weakly depends on the niobium concentration or the film thickness. As is usual for disordered materials, the relation between the applied power $P$ on a film of volume $\Omega$ and the electron and phonon temperatures is given by \cite{Marnieros2000}:
\begin{equation}
\frac{P}{\Omega}=g_{e-ph}\left(T_e^5-T_{ph}^5\right)
\end{equation}
\noindent with $g_{e-ph}$ the electron-phonon coupling constant. The thermal conductivity is then given by $G_{e-ph}=\frac{\partial P}{\partial T_e}V$ where $V$ is the thin film volume. The response time $\tau_{e-ph}=\frac{C_e}{G_{e-ph}}$ then has been evaluated considering the relevant temperature dependance of the specific heat $C_e$ \cite{Marnieros2000}. Typical results are given in table \ref{tab:2} for the 175 {\AA} thick Nb$_{14}$Si$_{86}$ film at 50 and 100 mK. At 100 mK, this film is still close to mid-transition and has both a NEP suitable for terrestrial telescope applications and a sufficiently short response time. The NEP can be improved down to a few 10$^{-18}$ $W/\sqrt{Hz}$ by lowering the operating temperature to 50 mK, however at the cost of a slightly longer - but still acceptable - response time. These films thus have performances comparable to currently used state-of-the-art bolometers and present the advantage of a relatively simple fabrication process. They indeed can be synthesized on a regular substrate and do not require the use of membranes. This considerably simplifies the fabrication of bolometer matrices with a large number of pixels.

\begin{table}
\caption{Thermal properties of the 175 {\AA} thick Nb$_{14}$Si$_{86}$ sample : electron-phonon coupling constant $g_{e-ph}$, thermal conductivity $G_{e-ph}$, estimated electron-phonon response time $\tau_{e-ph}$ and NEP.}
\label{tab:2}
\begin{tabular}{lllll}
\hline\noalign{\smallskip}
T (mK)& $g_{e-ph}$ (W.K$^{-5}$.cm$^{-3}$) & $G_{e-ph}$ (W.K$^{-1}$) & $\tau_{e-ph}$ (ms) & NEP ($W.\sqrt{Hz}$)\\
\noalign{\smallskip}\hline\noalign{\smallskip}
50 & 26.8  &  10$^{-9}$ & 0.97 & 2.3$\times10^{-17}$\\
100 & 41.8  & 9.8$\times10^{-11}$ & 8.4 & 3.7$\times10^{-18}$\\
\noalign{\smallskip}\hline
\end{tabular}
\end{table}

\section{Future prospects}
\label{sec:Future_prospects}

\begin{figure}
\includegraphics[width=0.8\textwidth]{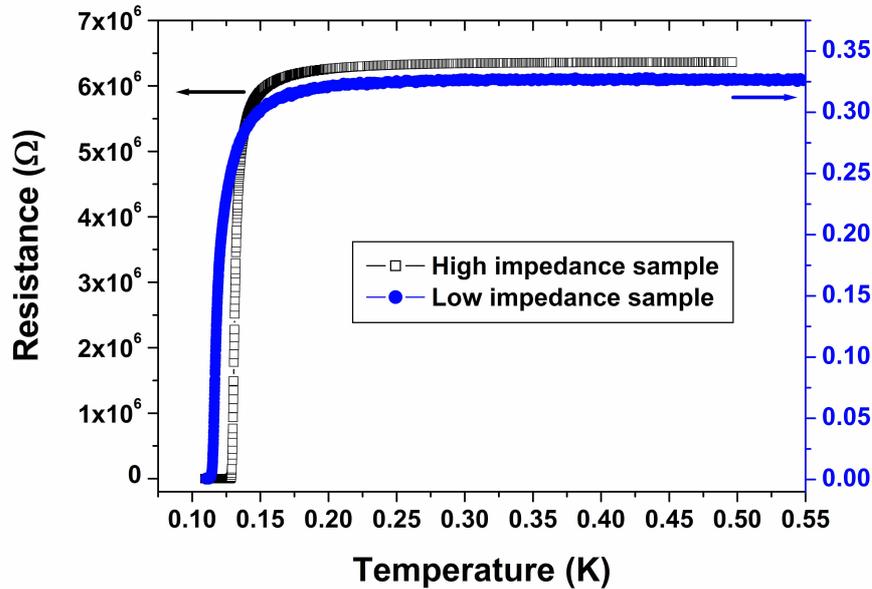}
\caption{Resistance as a function of temperature for superconducting films which design has been adapted for either JFET or SQUID readout.}
\label{fig:2}
\end{figure}

Regarding possible applications for radiation detection, the tunable superconducting properties of amorphous Nb$_x$Si$_{1-x}$ thin films allow the tuning of the operating temperature while maintaining an optimal incoming wave absorption. Moreover, the low electron-phonon coupling in these films enable us to take advantage from the electron-phonon decoupling to build an efficient bolometer with state-of-the-art NEP and which does not require elaborate fabrication process. Another concern for these bolometer matrices fabrication is the adaptation of the pixel to the readout electronics. Indeed, if thousands of pixels are to be implemented, the readout multiplexing is an important technological issue. Currently, the TES are mainly measured by a SQUID-based electronics. However a JFET-based electronics could be an interesting alternative.

An notable feature of the a-Nb$_x$Si$_{1-x}$ films concerns their adaptability to both considered electronic readouts. Indeed, the important parameter for radiation absorption is the sheet resistance. However, the films can be designed to yield a high resistance sample (typically a few M$\Omega$) suitable for JFET readout through a large Length/Width ratio. On the contrary, a small value of such a ratio will give a low resistance (typically less than 1 $\Omega$) sample that could be probed by a SQUID readout. We have fabricated two such geometries (figure \ref{fig:2}), based on 50 nm-thick a-Nb$_{14}$Si$_{86}$ films which superconducting temperatures have been tuned around 100 mK by a specific annealing at $120\,^{\circ}\mathrm{C}$.

The flexibility of these a-Nb$_x$Si$_{1-x}$ superconducting films and the versatility of the readout systems that could be used make them particularly interesting for applications in detection that could extend beyond CMB measurements.




\end{document}